\documentclass{jaa}
\usepackage[para,online,flushleft]{threeparttable}
\usepackage[numbers]{natbib}
\usepackage[pdftex]{graphicx}
\usepackage{adjustbox}
\usepackage{float}
 \usepackage{grffile}
 
\begin{document}\sloppy

\title{Exploring Sub-MeV Sensitivity of AstroSat$-$CZTI for ON$-$axis Bright Sources }

\author{Abhay Kumar\textsuperscript{1,2,*}, Tanmoy Chattopadhyay\textsuperscript{3,*}, Santosh V Vadawale\textsuperscript{1}, A.R. Rao\textsuperscript{4,5}, Soumya Gupta\textsuperscript{4}, Mithun N. P. S.\textsuperscript{1}, Varun Bhalerao\textsuperscript{6} and Dipankar Bhattacharya\textsuperscript{4}}

\affilOne{\textsuperscript{1}Physical Research Laboratory, Navrangpura, Ahmedabad, 380009, India.\\}

\affilTwo{\textsuperscript{2}Indian Institute of Technology, Gandhinagar, 382355, India.\\}

\affilThree{\textsuperscript{3}Kavli Institute of Astrophysics and Cosmology, 452 Lomita Mall, Stanford, CA 94305, USA\\}
\affilFour{\textsuperscript{4}The Inter-University Centre for Astronomy and Astrophysics, Pune, India\\}
\affilFive{\textsuperscript{5}Tata Institute of Fundamental Research, Mumbai, India\\}
\affilSix{\textsuperscript{6}Indian Institute of Technology Bombay, Mumbai, India\\}

\twocolumn[{

\maketitle

\corres{abhaykk@prl.res.in, tanmoyc@stanford.edu}

\msinfo{--}{--}

\begin{abstract}
The Cadmium Zinc Telluride Imager (CZTI) onboard {\em AstroSat} is designed for hard X-ray imaging and spectroscopy in the energy range of 20 - 100 keV. The CZT detectors are of 5 mm thickness and hence have good efficiency for Compton interactions beyond 100 keV. The polarisation analysis using CZTI relies on such Compton events and have been verified experimentally. The same Compton events can also be used to extend the spectroscopy up to 380 keV. Further, it has been observed that about 20$\%$ pixels of the CZTI detector plane have low gain, and they are excluded from the primary spectroscopy. If these pixels are included, then the spectroscopic capability of CZTI can be extended up to 500 keV and further up to 700 keV with a better gain calibration in the future. Here we explore the possibility of using the Compton events as well as the low gain pixels to extend the spectroscopic energy range of CZTI for  ON-axis bright X-ray sources. We demonstrate this technique using Crab observations and explore its sensitivity.
\\

\end{abstract}

\keywords{{\em AstroSat}---CZT-Imager---sub-Mev spectroscopy---Crab.}
}]

\doinum{12.3456/s78910-011-012-3}
\artcitid{\#\#\#\#}
\volnum{000}
\year{0000}
\pgrange{1--}
\setcounter{page}{1}
\lp{1}

\section{Introduction}

The Cadmium Zinc Telluride Imager (hereafter CZTI) onboard {\em AstroSat}, India's first dedicated astronomy satellite (Paul 2013; Singh \emph{et al}. 2014), is primarily designed for hard X-ray coded mask imaging and spectroscopy in the energy range of 20 - 100 keV  (Bhalerao \emph{et al}. 2017).  It consists of 64 Cadmium Zinc Telluride (CZT) detector modules having 5 mm thickness and 4 cm $\times$ 4 cm in dimension. Each module is further segmented spatially in 16 $\times$ 16 array of pixels (2.5 mm $\times$ 2.5 mm of pixel size). 
After the launch of  {\em AstroSat}, about 20$\%$ of the CZTI pixels (therefore $\sim$ 20\% of the total 924 $cm^2$ geometric area) were found to have a relatively lower gain than that of the spectroscopically good pixels, which makes these pixels sensitive to higher energy photons ($\sim$ 70 - 1000 keV for a gain shift $>$4). The 5 mm thick CZT detector provides sufficient detection efficiency up to 1 MeV. Motivated by this, we tried to explore the possibility of including these low gain pixels in the analysis to enhance the spectroscopic sensitivity of CZTI up to the sub-MeV region.

\par
 The Coded mask spectrum generated using the standard pipeline is restricted up to 100 keV, where the background is simultaneously obtained from the coded mask imaging. Above 100 keV, the 0.5 mm thick tantalum mask becomes increasingly transparent, along with the collimators and CZTI support structures. To obtain the high energy spectra above 100 keV in the absence of simultaneous background measurements requires a careful selection of blank sky observations. 
 Background flux depends on multiple factors like the spacecraft's geometric location in orbit, orbital precession of the satellite, and the time spent within the high background  South Atlantic Anomaly (SAA) region in an orbit. These contribute to a systematic modulation in the flux along the satellite's orbit, hence making the background subtraction quite challenging.

 Another challenge is to calibrate these pixels in the absence of any mono energetic lines at energies above 100 keV to estimate their gains. A careful calibration of these pixels have been attempted in a companion paper by Chattopadhyay \emph{et al}. 2020 in this issue, and spectroscopy up to 900 keV is explored for  Gamma Ray Bursts (GRBs). Compared to the ON-axis sources, spectroscopy of GRBs is relatively easy because of the availability of background (from the immediate pre-GRB and post-GRB observations) and the significantly higher signal to noise ratio for the GRBs. On the other hand, the ON-axis bright astrophysical sources are fainter, and hence longer exposure observations are required for sufficient detection,  which leads to more instrumental, charged particle, and cosmic X-ray background contributions.

This paper outlines the methodology of sub-MeV spectroscopy with CZTI for bright ON-axis sources. We utilize the 2-pixel Compton events, which are used to extract polarimetry information of the X-ray photons (Chattopadhyay \emph{et al}. 2014, 2019; Vadawale \emph{et al}. 2015, 2018) to enhance the spectroscopic capability. The selection of background observations and the background subtraction used for polarimetric measurements is described in detail here, with an emphasis on spectroscopy. We utilize the low gain pixels to extend the spectroscopic energy range of the instrument well beyond the standard limit.
 With the inclusion of the low gain pixels, Compton spectroscopy was also extended to 500 keV. We use the {\em AstroSat} mass model in GEANT4 (Agostinelli \emph{et al}. 2003) to generate the spectral response.

Here we carry out broadband spectroscopy of Crab using standard spectroscopic events (30 - 100 keV), 2-pixel Compton events including low gain pixels (100 - 500 keV), and also explored the 1-pixel events including low gain pixels (100 - 700 keV) to establish the sub-MeV spectroscopy methods and at the same time try to constrain the spectral parameters of Crab at higher energies.
The X-ray emission coming from  Crab Nebula can be divided into three parts: pulsed point source from the neutron star, synchrotron emission powered by charged particles coming from the centrally located pulsar, and the large diffused emission region in the Nebula. Toor $\&$ Seward (1974) fitted the Crab spectra in 2 - 60 keV using a power law and concluded that it could be used as a standard calibration source due to its steady nature. It is believed that it appears to be steady on a time scale of a few years because most of the emission is from the diffused expanding ejecta, which is extended. There is, however, no theoretical basis that the pulsed emission will be steady.  Different models have been tried to explain the stable behavior and emission mechanism of  Crab.  It is often described by a power law model (Kirsch \emph{et al}. 2005; Kuiper \emph{et al}. 2001). A broken power law is also used to describe the slope evolution, with a break around 100 keV (Strickman \emph{et al}. 1979; Ling $\&$ Wheaton 2003). Massaro \emph{et al}. (2000) and Mineo \emph{et al}. (2006)  used a single curved power law with a variation of the slope with Log(E) to describe the spectra. The {\em INTEGRAL/SPI} data were fit with a broken power law (Jourdain $\&$ Roques  2008) and also by the  Band model generally used for GRBs (Band \emph{et al}. 1993; Jourdain $\&$ Roques 2020). There is no general agreement on the best overall spectral model so far. Further, the absolute flux, emission mechanism, and the cause of Crab's stability are also not well known. The extended bandwidth of CZTI  using the Compton and single spectrum, including low gain pixel events, can help us understand the spectra of the Crab. It is to be noted that CZTI is also sensitive to polarisation, and hence a  simultaneous measurement of polarisation along with spectroscopy can help us in a better understanding of the Crab emission mechanism and the cause of the stability in its emitted flux.

In section \ref{sec_observation}, a brief description of the observations and analysis procedure is given. The results obtained are presented in section \ref{result}. Finally, in section \ref{discussion}, we discuss the sub-MeV sensitivity of CZTI and future plans.

\section{Observations and analysis procedure}\label{sec_observation}

There are many observations of Crab over the past five years of operation of {\em AstroSat}. We have selected those observations with sufficient exposure and also having a suitable background observation.
 The selection of appropriate background observation and subtraction is an important part of the analysis. The CZTI support structure becomes increasingly transparent above 100 keV. Therefore, background measurements can be affected due to the presence of bright X-ray sources within  70${^\circ}$ of the pointing direction of  CZTI. The Crab and Cygnus X-1 are two bright sources which should be avoided during the background observation. Since both the sources are located almost opposite to each other in the sky, it is possible to find a good region, away from these two sources, for the background observations. The background observation should also be close to the source observation time to avoid the error in some long-term secular variations in the background behavior.  There are a few such observations that satisfy the criterion of the background selection. Based on these considerations, we have finally selected five observations between 2015 and 2017 for further analysis. Details of the Crab observations and the corresponding background observations selected for the present analysis are given in Table \ref{observations}.

\begin{table*}
	
	\caption{Summary of Crab and blank sky observations.}	
	\label{observations}
	\begin{threeparttable}	
		\centering
		\renewcommand{\arraystretch}{1.5}
		\resizebox{\textwidth}{!}{%

	\begin{tabular}{p{2cm}|p{2.2cm}|p{1.5cm}|p{2cm}|p{2.2cm}|p{1.5cm}|p{1cm}|p{1cm}}
	\hline
    \multicolumn{3}{c}{Crab} & \multicolumn{5}{|c}{Blank Sky} \\
		\hline
	
		 ObsID & Date & Exposure  & ObsID& Date & Exposure & RA & DEC \\
		 &(yyyy/mm/dd) &(ks) & & (yyyy/mm/dd) & (ks) & (deg)& (deg)\\
     	\hline
		  $9000000096$ & 2015/11/12  &$ \, \rm$ 41 & $9000000276$ &
		  2016/01/16 & 64 & 183.48 & 22.8 \\[0.2cm]
		  
		  $9000000252$& 2016/01/07  & $\, \rm$ 60 & $9000000276 $& 2016/01/16 & 64 & 183.48 &22.8   \\[0.2cm]
		  
		  $9000000406$ & 2016/03/31  &$\, \rm$  114& $9000000404$ & 2016/03/29 & 64 & 228.21 & -9.09 \\[0.2cm]
		  
		  $9000000964$ & 2017/01/14  & $\, \rm$ 78& $9000000974$& 2017/01/22 & 51 & 183.48 & 22.8   \\[0.2cm]
		 
		  $9000000970$ & 2017/01/18  & $\, \rm $ 123&   $9000000974$ & 2017/01/22 & 51 & 183.48 & 22.8 \\[0.2cm]
		
		\hline
\hline

\end{tabular}}

\begin{tablenotes}
\small
\item Crab RA: $83.63^{\circ}$ and DEC: $22.01^{\circ}$
\end{tablenotes}
\end{threeparttable}
\end{table*}

\subsection{Single and Compton event selection}{\label{event_sel}}
CZTI is a pixelated detector. The scientific data analysis of CZTI is done with two types of CZTI events: 1-pixel or single pixel events and 2-pixel events. The single pixel events registered in the CZTI are considered as the true 1-pixel events if there is no event registered within 100 $\mu$s time window on either side of the single event. The events in the CZTI are time stamped at every 20 $\mu$s (Bhalerao \emph{et al}. 2017) and any two events occurring within 20 $\mu$s coincidence time window in two pixels are considered as true 2-pixel event (Chattopadhyay \emph{et al}. 2014). The standard single pixel mask-weighted spectra in 30 - 100 keV (hereafter EB1) is generated following standard pipeline software available at the AstroSat science support cell (ASSC) \footnote{http://astrosat-ssc.iucaa.in/}.

After the launch of {\em AstroSat}, it is observed that CZTI had low gain pixels of about 20$\%$ of the detector plane.  
These pixels are found to have a shift in gain (gs) by a factor of 1.5$-$5 compared to that calculated during the ground calibration such that a mono-energetic line of energy `E' now appears at a lower energy PI bin (E/gs); hence the name low gain pixels. Because these pixels now record higher energies, the spectroscopic range of CZTI with single pixel events can be extended up to 700 keV with the inclusion of low gain pixels. Detailed characterisation methods of low gain pixels have been discussed in the companion paper by Chattopadhyay \emph{et al}. 2020. Above 100 keV single event spectra, including the low gain pixel events, need to be analysed outside of the standard pipeline because they are not included in the standard pipeline of CZTI. The initial processing is to remove the intervals of high background during the passage of South Atlantic Anomaly from each observation and removal of the noisy (pixel having counts more than five sigma above mean value) or spectroscopically bad pixels (energy resolution is poor compared to the  normal pixels). This procedure is valid for both the single and 2-pixel event spectra. After selecting the single pixel events, energy deposited in each pixel is used to generate the single pixel spectrum including low gain pixels in the 100 - 700 keV (hereafter EB2). We have binned the events into 60 channels with 10 keV energy bin sizes in the 100 - 700 keV energy range. 

\par
Above 100 keV, the 5 mm thick CZTs have sufficient efficiency for Compton interactions.  Polarisation analysis in the 100 - 380 keV range depends upon such Compton events  (Chattopadhyay \emph{et al}. 2014; Vadawale \emph{et al}. 2015). These  Compton events are used here to do spectroscopy in the 100 - 380 keV range. After incorporating the low gain pixels, Compton spectroscopy can be further extended to 500 keV. For the Compton event selection, we follow the 2-pixel Compton event selection criteria, as discussed in Chattopadhyay \emph{et al}. 2014. The readout logic in the CZTI is such that it reads events from one module at a time. If two events are registered in two different pixels in the same module, then it is possible that two events would get two different time stamps. Hence, all the events occurring within the coincidence time window of 40 $\mu$s are selected for the analysis. These events are further filtered through Compton kinematics criteria to enhance the signal to noise ratio. 
After the selection of the Compton events, the sum of the energy deposited in the scattering and the absorption pixel is used to generate the 2-pixel Compton events spectrum including low gain pixels in the 100 - 500 keV (hereafter EB3). We have binned the data at 10 keV energy bin size in the 40 channels ranging from 100 keV to 500 keV.

\subsection{Background subtraction: Phase match method}\label{back_sub}
The primary sources of background in the CZT detector are the Cosmic X-ray background, the Earth's X-ray Albedo, and the locally produced X-rays due to Cosmic ray interactions.  The Compton background in the detector is due to the Compton scattering of these background X-rays.   In addition to this, a small part of the Compton background consists of chance coincidence events within 40 $\mu$s coincidence time window. \par

\begin{figure}[!ht]
	\centering
	
	\includegraphics[scale=0.47]{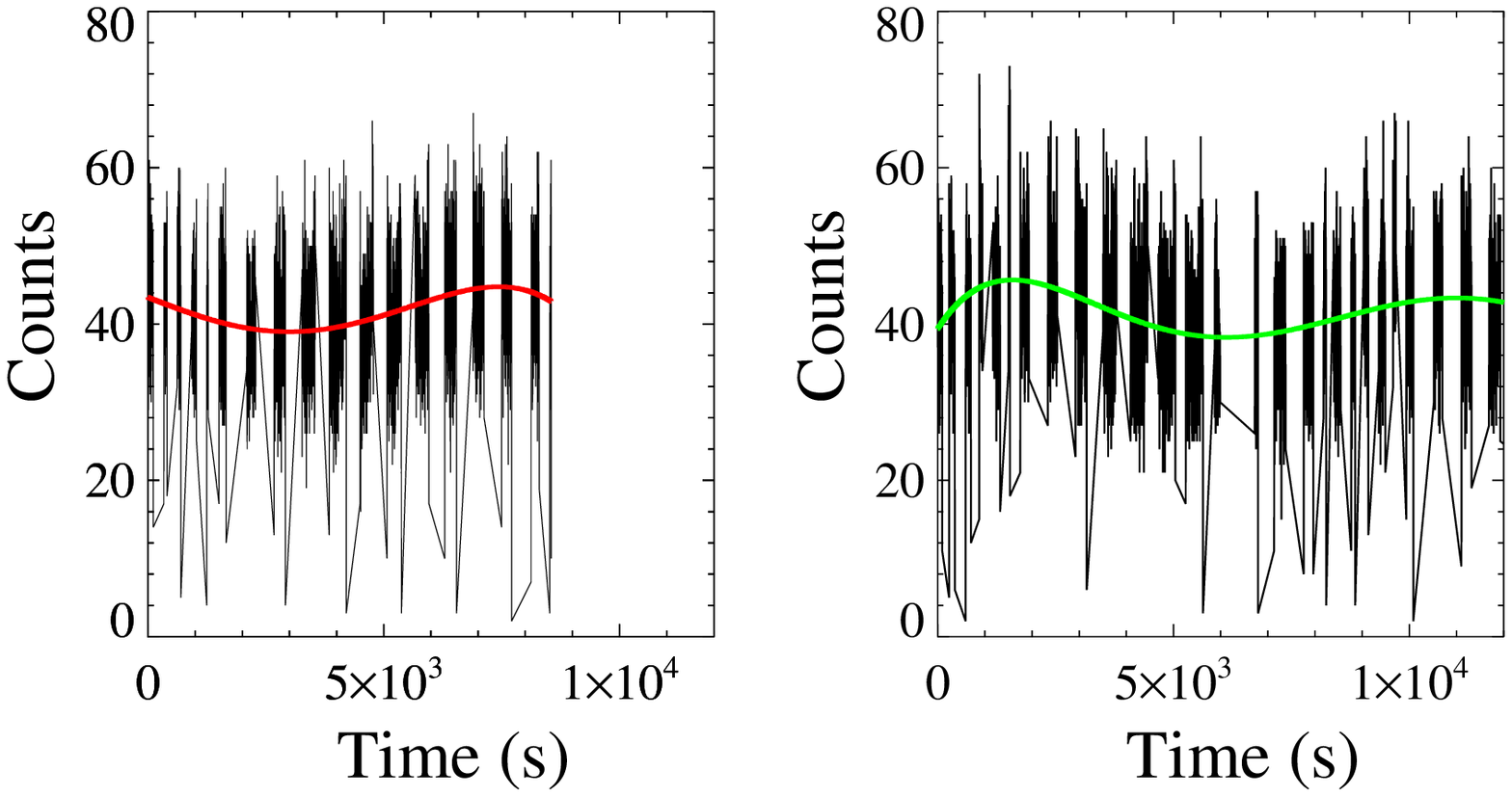}

	\includegraphics[scale=0.47]{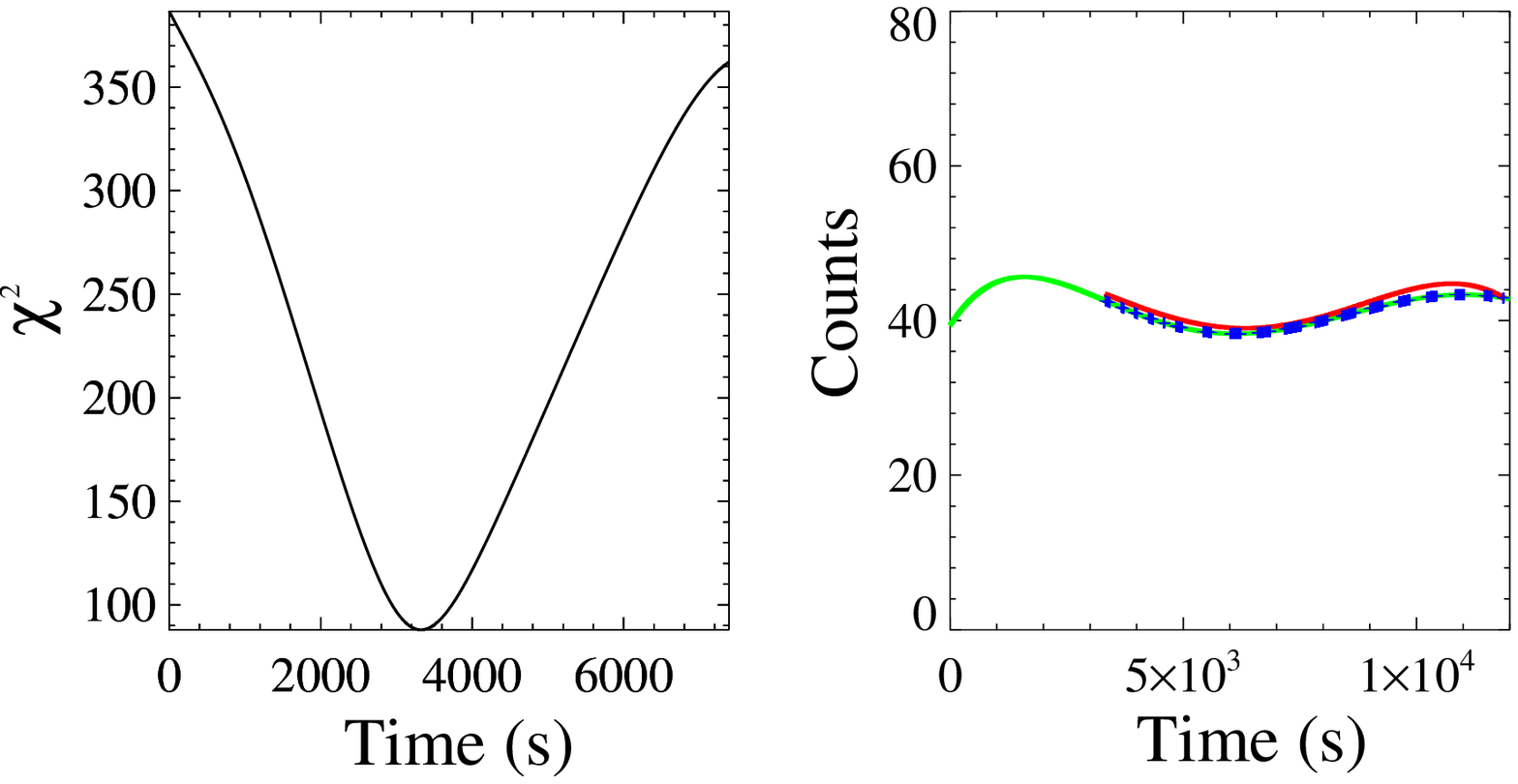}
	\caption{Top panel  shows the light curves of Crab (ObsId 96) on the left  and background (ObsId 276) on the right, both   fitted with  fifth degree polynomials (red curve for Crab and green curve for background). Bottom left figure shows the variation of $\chi^2$ as a function of shift in the Crab light curve with respect to the background, to determine the time for `phase matching'. Bottom right figure shows the `phase matched' background and Crab polynomials, in the  green curve and red, respectively. The blue curve represents the background where Crab observation is available.} 
	\label{phasematch96}
\end{figure}

The background events from the blank sky observations are filtered through the same selection criteria as the source, as discussed in section \ref{event_sel}. The observed background counts show a prominent orbital variation as well as a diurnal variation depending on the geometric location of the spacecraft in the orbit, orbital precession, and the location and duration of the SAA region in orbit. 
All these contribute to a systematic modulation (see the top panel of Figure \ref{phasematch96}) in the flux along the orbit of the satellite within the duration of the observation. Because of the modulating flux, it is important to select similar portions of the Crab and background orbits based on the spacecraft's ground tracks (latitude and longitude) and use them for further analysis. But this puts a strict condition on the selection of the orbits and leaves a short usable exposure of Crab and background observations. An alternate method developed for background subtraction is to match the phase of the background and Crab light curves (the `phase match method'), used for the polarisation measurements of Crab (Vadawale \emph{et al}. 2018). In the phase match method, first, the Crab and background light curves are fitted with an appropriate higher-order (here 5$^{th}$ order) polynomial. Then the  Crab and background light curves are matched by sliding the Crab polynomial over the background polynomial every 10 seconds and estimating the best match by minimising the  $\chi^2$. Within the matched region of the two polynomials, the background is taken only for those time regions for which there is a source observation (see bottom right of Figure \ref{phasematch96}). From the background's phase-matched region,  we calculate the ratio of the average background count rate to the count rate in the phase matched region (`correction factor') and multiply that to the total background exposure to calculate an effective background exposure. The use of effective exposure automatically takes care of the different phases of source and background observations. For example, if the source and backgrounds are observed in the same phase, the multiplication factor will be close to 1. For longer source observations (exposure $\gg$ background exposure), it is divided into multiple parts 
and for each part of the source observations, we calculate the correction factor in the way described above and then the final correction factor is calculated as the weighted average. 

\begin{figure}[!ht]
	\centering
	\includegraphics[scale=0.15]{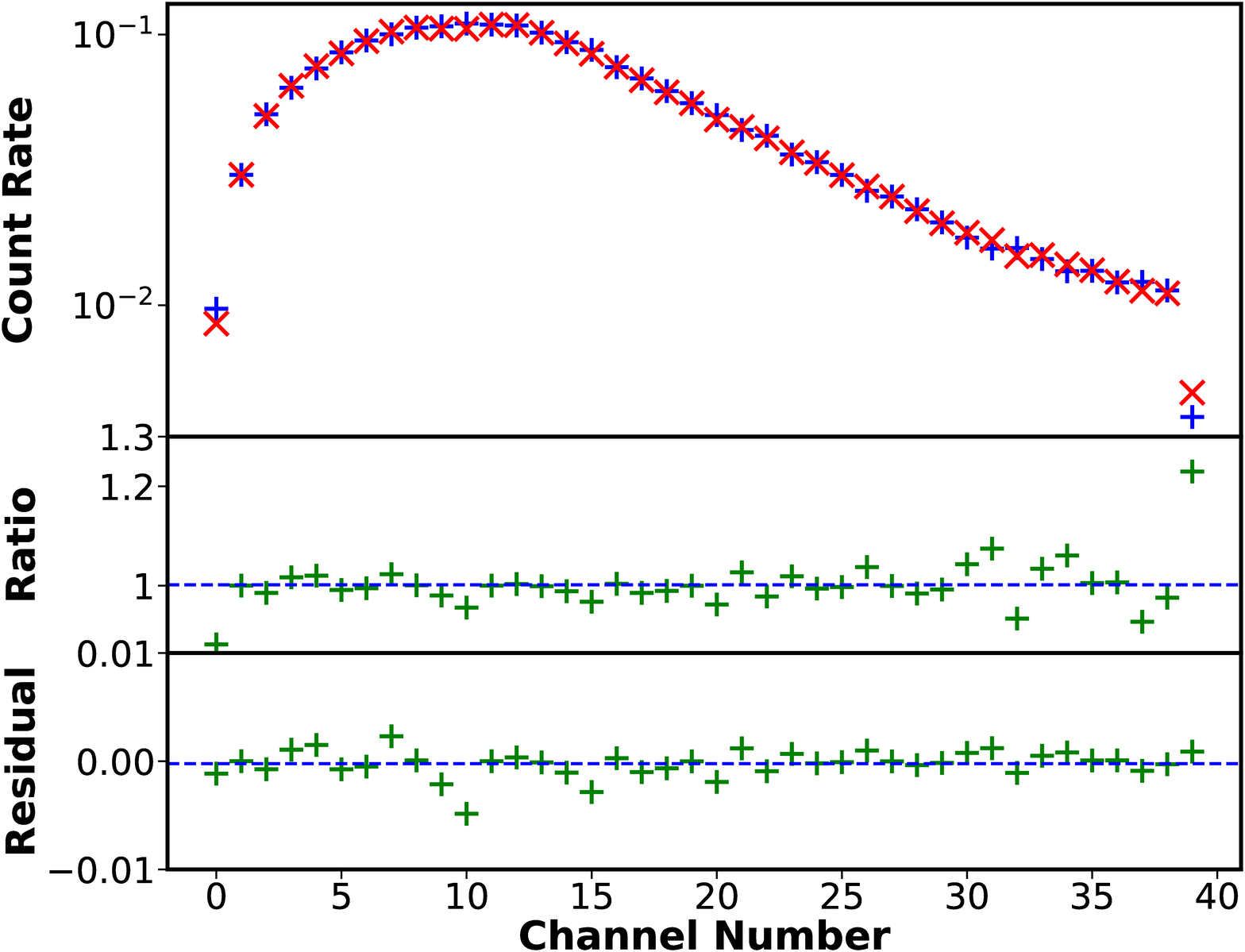}
	
	\includegraphics[scale=0.15]{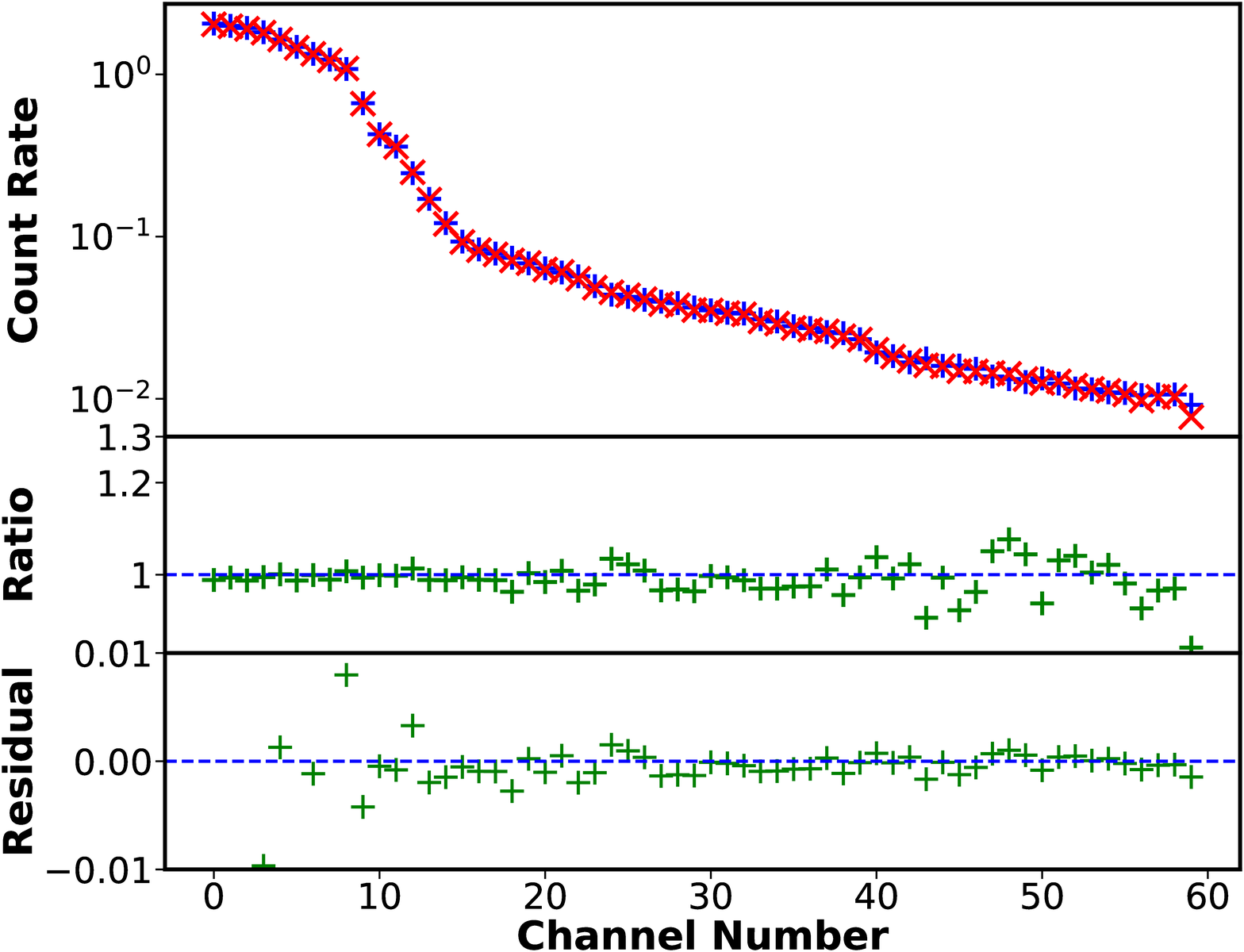}
	
	\caption{ The top figure shows the phase matched 2-event Compton spectra  (EB3)  of the blank sky observation CU (red) and background B (blue) from 100 keV to 500 keV. The ratio of the two spectra is shown in the middle panel and the  residuals of the two are shown in the bottom panel. The bottom figure shows similar plots for the low gain pixel spectra (EB2) from 100 keV to 700 keV}. 
	\label{phasematch_CU_B}
\end{figure}

It is to be noted that the phase match method only ensures that the source (Crab in this case) and blank sky background data are taken for the same orbital phase (spacecraft orbit). It is not for the background spectral modeling. The background scaling is done to the phase matched background region to best mimic the actual background during the source observation. Because the background spectra are obtained from observations of “blank sky” with no other hard X-ray sources in the FOV (confirmed from BAT catalog), the spectra for the used backgrounds are expected to be the same. 
To demonstrate this, we selected a CZTI observation from UVIT catalog (ObsID 1008) such that there are no other bright X-ray sources in the FOV of CZTI (call this ‘CU’). We then used the polynomial method to do phase match between CU and background data ObsID 974 (call this ‘B’), which is used in the analysis and correct for total flux in B for the phase of CU. Because CU is essentially a blank sky background for CZTI, similar flux and spectra for both B and CU after phase correction is expected. 

We found identical flux for both B and CU (see Figure \ref{phasematch_CU_B}), signifying that the polynomial method is capable of finding the common phase and scale the flux accordingly. The spectra are also found to be identical, justifying the underlying assumption that the blank sky observations for CZTI with the predefined selection criteria yield a similar photon energy distribution.

\subsection{Spectral Response}\label{response}

The response for  EB1 is generated using the standard pipeline of the {\em AstroSat} and for both EB2 and the EB3 using the GEANT4 simulation of the {\em AstroSat} mass model. Details of the {\em AstroSat} mass model and it's validation has been discussed in Chattopadhyay \emph{et al}. 2019 and Mate \emph{et al}. 2020, in this issue. We simulated the mass model for 56 mono-energies ranging from 100 keV to 2 MeV  (at every 20 keV up to 1 MeV and 200 keV till 2 MeV) for $10^9$ photons. For each energy, the distribution of deposited energy in CZTI pixels is computed at 1 keV binning for each pixel. The CZTI pixel-level LLD (Lower Level Discriminator), the ULDs (Upper Level Discriminator), list of noisy and dead pixels obtained from the actual observational data are applied to the simulation data. For 2-pixel Compton events, the sums of the energies in the corresponding two pixels are used to obtain the total deposited energy, while for single event response, total absorbed energy for a given incident photon is used. We applied the same criterion of event selection, as discussed in section \ref{event_sel}.
We then convolve the 1 keV bins by a Gaussian of 8 keV Full Width at Half Maximum (FWHM). We have not noticed any significant increase in FWHM with energy for CZTI pixels during ground calibration. Therefore, FWHM is kept constant across the energy. It is to be noted that the response for EB1 is generated using  $\mu \tau$ and charge diffusion based line profile model (Chattopadhyay \emph{et al}. 2016).

\begin{table*}
	
	 \resizebox{0.90\textwidth}{!}
	 {\begin{minipage}{\textwidth}
	\caption{Comparison of fitted parameters between {\em INTEGRAL/SPI} and {\em AstroSat} for broken power law. The errors are reported for $90\%$ confidence interval. The mean value of the parameters of all the observations for different combinations of data is mentioned in the bottom row of each block.}
	\label{integral_czti_results}
	\begin{tabular}{*{12}{l}}

		\hline
		&Instrument&Spectra&ObsID & $PhoIndx1$ & $PhoIndx2$ &$E_{break}$& $Norm$ & $Flux(30 - 100 keV)$  & $\chi^2$/dof\\
		 & & & &  & &(keV) &&$10^{-8}$(erg/cm$^2$s)&   \\
		\hline
		 & INTEGRAL& & &$2.08^{+0.01}_{-0.01}$  &  $2.23^{+0.05}_{-0.05}$ &$100^{*}$ &$9.3^{+0.14}_{-0.14}$ &$1.30$  &  \\[0.2cm]  
		 
		  & CZTI & $EB1\&EB3$& 96 & $2.18^{+0.05}_{-0.05}$ & $1.80^{+0.48}_{-0.47}$ &  $100^{*}$ &$5.98^{+1.27}_{-1.04}$&$ {0.56^{+0.10}_{-0.09}}$& 232/175 \\[0.2cm] 
		 
		 & & & 252 & $2.08^{+0.03}_{-0.03}$ & $1.72^{+0.44}_{-0.48}$ &  $100^{*}$ &$6.00^{+0.84}_{-0.73}$&${0.85^{+0.12}_{-0.10}}$& 238/172 \\[0.2cm]  
		 
		 & & & 406 & $2.12^{+0.03}_{-0.03}$ & $1.63^{+0.39}_{-0.36}$ &  $100^{*}$ &$7.5^{+0.78}_{-0.70}$&${0.89^{+0.09}_{-0.08}}$ & 288/175 \\[0.2cm] 
		 
		 & & & 964 & $2.08^{+0.04}_{-0.04}$ & $2.65^{+0.90}_{-0.74}$ &  $100^{*}$ &$5.73^{+0.90}_{-0.77}$&${0.82^{+0.11}_{-0.12}}$& 244/172 \\[0.2cm]

		 & & & 970 & $2.10^{+0.03}_{-0.03}$ & $1.98^{+0.48}_{-0.44}$ &  $100^{*}$ &$6.44^{+0.72}_{-0.64}$&${0.82^{+0.09}_{-0.08}}$ & 328/172 \\[0.2cm]

		& & &{ mean }& ${2.11^{+0.02}_{-0.02}}$ & ${1.95^{+0.18}_{-0.18}}$ & ${100^{*}}$ & ${6.33^{+0.32}_{-0.32}}$& ${0.79^{+0.06}_{-0.06}}$ & &\\[0.2cm]

		 \hline
		 & & $EB1\&EB2$& 96 & $2.21^{+0.04}_{-0.04}$ & $1.20^{+0.24}_{-0.18}$ &  $100^{*}$ &$6.75^{+1.29}_{-1.09}$&${0.55^{+0.10}_{-0.09}}$ & 225/195 \\[0.2cm]

		 & & & 252 & $2.1^{+0.03}_{-0.03}$ & $1.33^{+0.20}_{-0.16}$ &  $100^{*}$ &$6.48^{+0.73}_{-0.66}$&${0.84^{+0.09}_{-0.09}}$ & 220/195 \\[0.2cm]

		 & & & 406 & $2.13^{+0.02}_{-0.02}$ & $1.46^{+0.17}_{-0.14}$ &  $100^{*}$ &$7.76^{+0.59}_{-0.55}$&${0.89^{+0.07}_{-0.06}}$ & 274/195 \\[0.2cm]

		& & & 964 & $2.13^{+0.03}_{-0.03}$ & $1.5^{+0.22}_{-0.17}$ &  $100^{*}$ &$6.98^{+0.69}_{-0.63}$&${0.80^{+0.08}_{-0.07}}$& 225/195 \\[0.2cm]

		 & & & 970 & $2.14^{+0.02}_{-0.02}$ & $1.34^{+0.17}_{-0.14}$ &  $100^{*}$ &$7.26^{+0.57}_{-0.54}$&${0.81^{+0.06}_{-0.06}}$ & 317/195 \\[0.2cm]

		 & & &{ mean }& ${2.14^{+0.02}_{-0.02}}$ & ${1.36^{+0.05}_{-0.05}}$ & ${100^{*}}$ & ${7.05^{+0.22}_{-0.22}}$& ${0.78^{+0.06}_{-0.06}}$ & &\\[0.2cm]

		 \hline
		 
		& & $EB1,EB2 \& EB3$& 96 & $2.21^{+0.04}_{-0.04}$ & $1.24^{+0.24}_{-0.18}$ &  $100^{*}$ &$6.68^{+1.28}_{-1.07}$& ${0.55^{+0.10}_{-0.09}}$ & 244/234 \\[0.2cm]

		 & & & 252 & $2.1^{+0.03}_{-0.03}$ & $1.36^{+0.2}_{-0.16}$ &  $100^{*}$ &$6.45^{+0.72}_{-0.66}$& ${0.84^{+0.09}_{-0.09}}$ & 321/234 \\[0.2cm]

		 & & & 406 & $2.13^{+0.02}_{-0.02}$ & $1.48^{+0.17}_{-0.14}$ &  $100^{*}$ &$7.71^{+0.58}_{-0.55}$& ${0.89^{+0.07}_{-0.06}}$ & 309/234 \\[0.2cm]

		 & & & 964 & $2.13^{+0.02}_{-0.02}$ & $1.5^{+0.23}_{-0.18}$ &  $100^{*}$ &$6.94^{+0.69}_{-0.64}$&${0.80^{+0.08}_{-0.07}}$   & 243/207 \\[0.2cm]

		 & & & 970 & $2.13^{+0.02}_{-0.02}$ & $1.38^{+0.18}_{-0.15}$ &  $100^{*}$ &$7.21^{+0.57}_{-0.54}$& ${0.81^{+0.06}_{-0.06}}$& 384/234 \\[0.2cm]

		  & & &{ mean }& ${2.14^{+0.02}_{-0.02}}$ & ${1.39^{+0.05}_{-0.05}}$ & ${100^{*}}$ & ${7.0^{+0.22}_{-0.22}}$& ${0.78^{+0.06}_{-0.06}}$& &\\[0.2cm]

		\hline
		
\hline
\end{tabular}

\tablenotes{{\em INTEGRAL/SPI} flux in the 30 - 100 keV is estimated using the model parameters given in (Jourdain $\&$ Roques 2008, 2020). $\ast$ is for fixed parameters.}

\end{minipage}}
\end{table*}

\section{Results }\label{result}

\subsection{Single and Compton event spectra (EB1 and EB3)\label{EB1_EB3_spectra}}

\begin{figure}[!ht]
	\centering
	\includegraphics[width=0.70\linewidth,angle=-90]{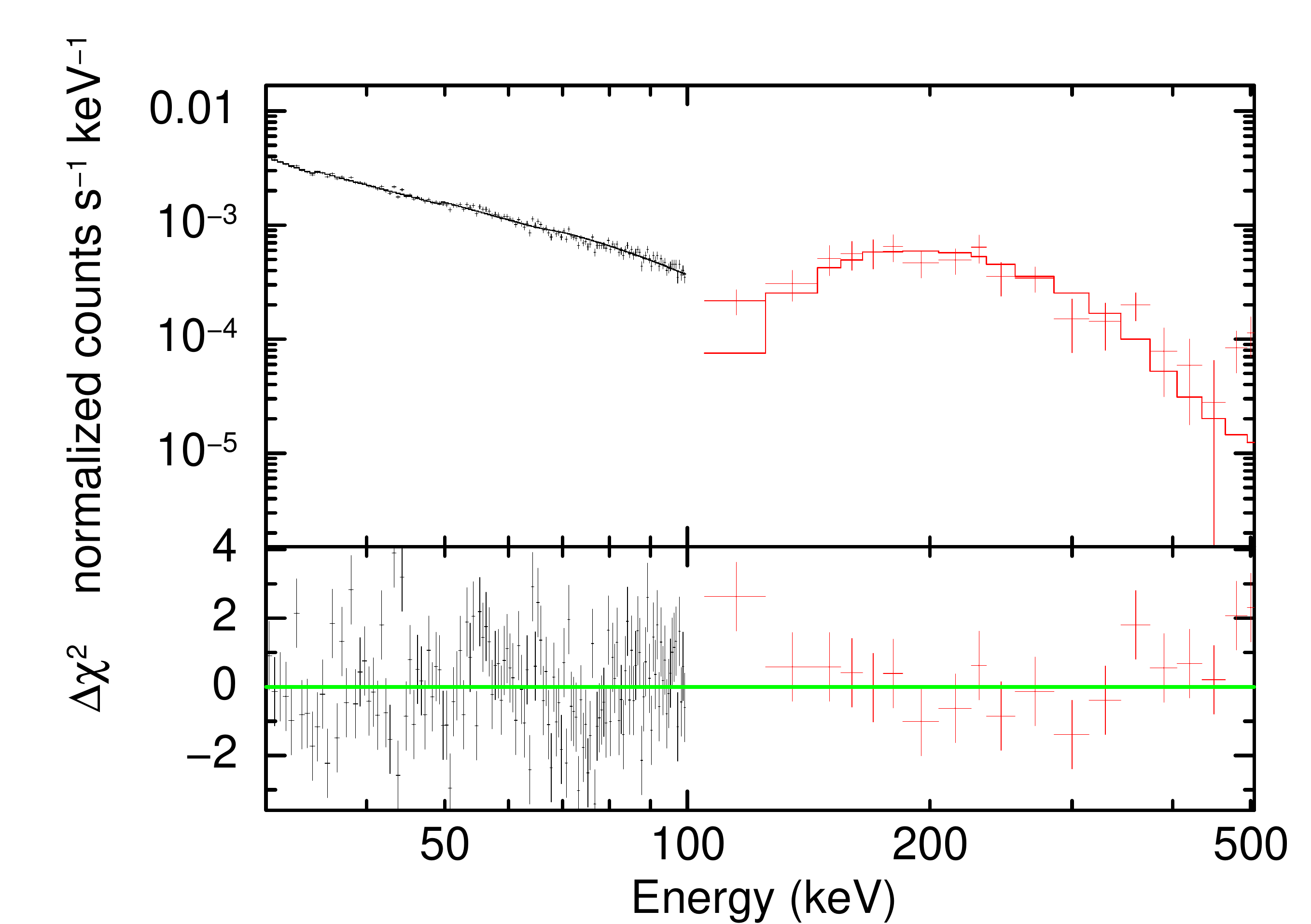}

	\caption{Broadband spectra of Crab (ObsID 406, 114 ks) fitted with a broken power law. The black, red colors are used for EB1, and EB3 respectively.} 
	\label{bknspec406_EB1_EB3}
\end{figure}

\begin{figure}[!ht]
	\centering

	\includegraphics[scale=.28]{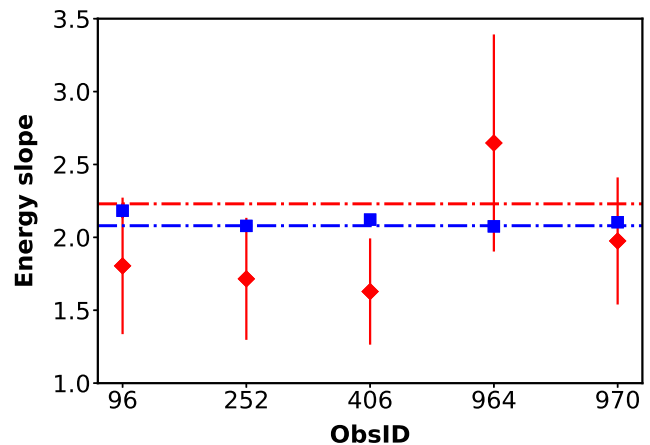}

	\caption{Best fit parameter of all the selected observations obtained after simultaneous fitting of EB1 and EB3 using broken power law model. Red diamonds represent higher energy slope ($PhoIndx2$) and the blue squares represent low energy slope ($PhoIndx1$). The dashed-dot horizontal lines represent the INTEGRAL value of $PhoIndx1$ (blue) and $PhoIndx2$ (red) in the corresponding energy regime.} 
	\label{bknpara}
\end{figure}

\begin{figure}[!ht]
	\centering
	\includegraphics[scale=0.6]{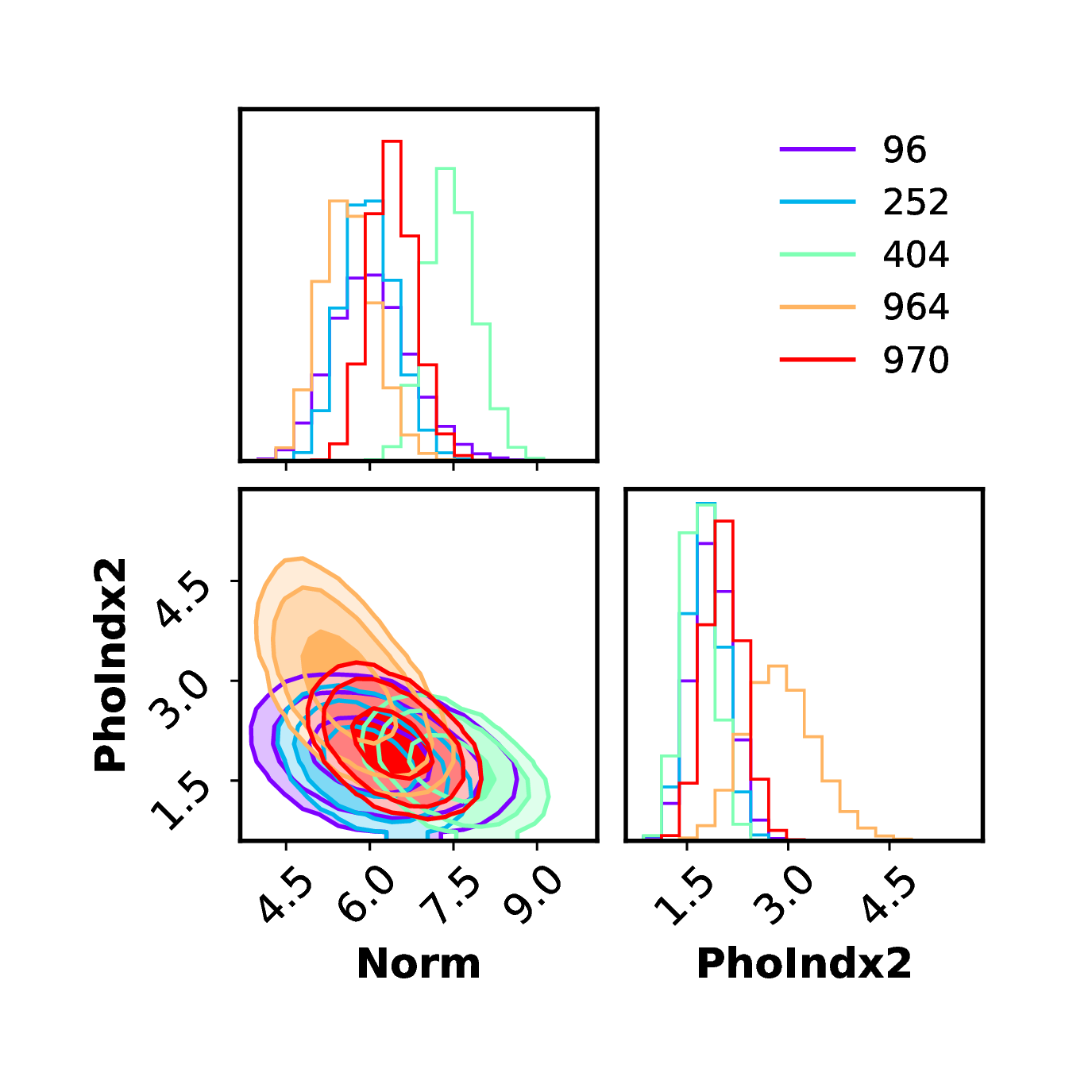}

	\caption{Corner plot of the $phoIndx2$ verses $Norm$ for all the five ObsIDs plotted together for EB1 and EB3. Different colors of contours represents different observations as shown in legend. The peak value of the $PhoIndx2$ is $2.04^{+0.48}_{-0.44}$ and the $Norm$  is $6.39^{+0.71}_{-0.63}$ } 
	\label{corner}
\end{figure}

We use the broken power law model to fit the Crab curved spectra in 30 - 500 keV. It has been long used to explain the Crab spectra with break energy at 100 keV (Strickman \emph{et al}. 1979; Ling $\&$ Wheaton 2003). {\em INTEGRAL/SPI} has also shown the spectral fitting using broken power law up to sub-MeV region (20 keV - 1 MeV) (Jourdain $\&$ Roques 2008). We have analysed all the selected observations and then fitted the resultant spectra (EB1 and EB3) simultaneously using const$\times$bknpower in {\tt XSPEC} (Arnaud 1996) freezing break energy at 100 keV while the other parameters (photon indices) are tied across the spectra. To account for the cross calibration and differences between the different spectra, a constant was multiplied to the model. It was fixed to one for EB1 and left free to vary for others. The EB1 below 30 keV and above 100 keV is ignored due to calibration issues.
No systematic has been added to the EB1 and EB3. Spectral fitting for one of the five  observations (ObsID 406, 114 ks) is shown in Figure \ref{bknspec406_EB1_EB3}. The values of the fitted parameters for all the five observations along with the {\em INTEGRAL/SPI} results are given   in Table \ref{integral_czti_results}. The low energy slope ($PhoIndx1$) and the higher energy slope ($PhoIndx2$) are well constrained and consistent  with the {\em INTEGRAL/SPI} (Jourdain E., Roques J. P. 2008) values within errors (see Figure \ref{bknpara}). The contour plots of $PhoIndx2$ versus $Norm$ for all the five observations are shown in Figure \ref{corner}. The inputs for the contour plot are generated using the chain command in {\tt XSPEC} after getting the best fit parameters, and the corner plots are generated using the python corner module (Foreman-Mackey 2017). The corner plots show that the  value of  $PhoIndx2$ is $2.04^{+0.48}_{-0.44}$ and that of $Norm$ is $6.39^{+0.71}_{-0.63}$, which is $\sim$ 31$\%$ smaller than the $Norm$ for {\em INTEGRAL/SPI}.

\begin{figure}[h!]
	\centering

	\includegraphics[scale=0.28]{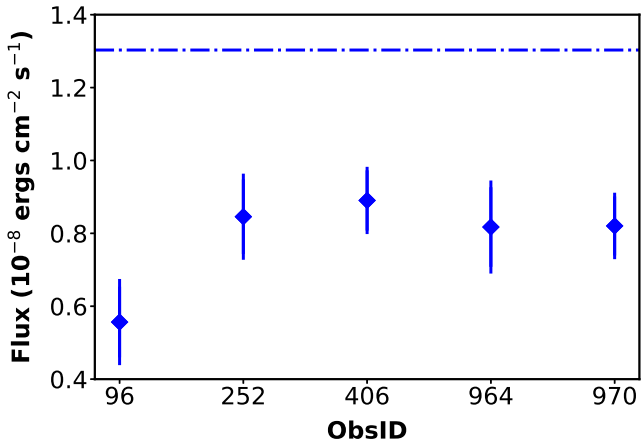}

	\caption{Estimated flux of the selected observations after fitting EB1 and EB3 simultaneously using broken power law . The blue diamond represents the CZTI value of the Crab flux. The dashed-dot horizontal line represent the INTEGRAL value of the Crab flux in the corresponding energy band.} 
	\label{bknflux}
\end{figure}

CZTI sensitivity is also studied by looking at the flux variation of Crab with time. Figure \ref{bknflux} compares the Crab flux covering the period between 2015 Nov 12 and 2017 Jan 18. Each data point corresponds to the flux obtained in each observation, and the dotted line represents the flux measured by  {\em INTEGRAL/SPI}. The flux from  ObsID 96 is much lower than the values found for the other ObsIDs. If we exclude the data from ObsID 96, the remaining four measurements are within 5\% of the mean value. We note here that the background for this  ObsID is measured more than two months before the Crab observation, whereas for all the other ObsIDs, the background is measured within ten days of the respective Crab observation. The effect of secular variations in the background on the flux measurements needs to be investigated further.

\subsection{Low Gain spectra}

As discussed earlier,  20$\%$ of the pixels in CZTI  are found to have lower gains as compared to the spectroscopically good pixels, which makes these pixels sensitive to higher energy photons ($\sim$ 70 - 1000 keV for a gain shift $>$4). Initially, these pixels were considered as bad pixels and removed from the analysis. Here, we have explored the spectroscopic capability of CZTI up to 700 keV, including these pixels.

\begin{figure}[!ht]
	\centering

	\includegraphics[width=0.72\linewidth,angle=-90]{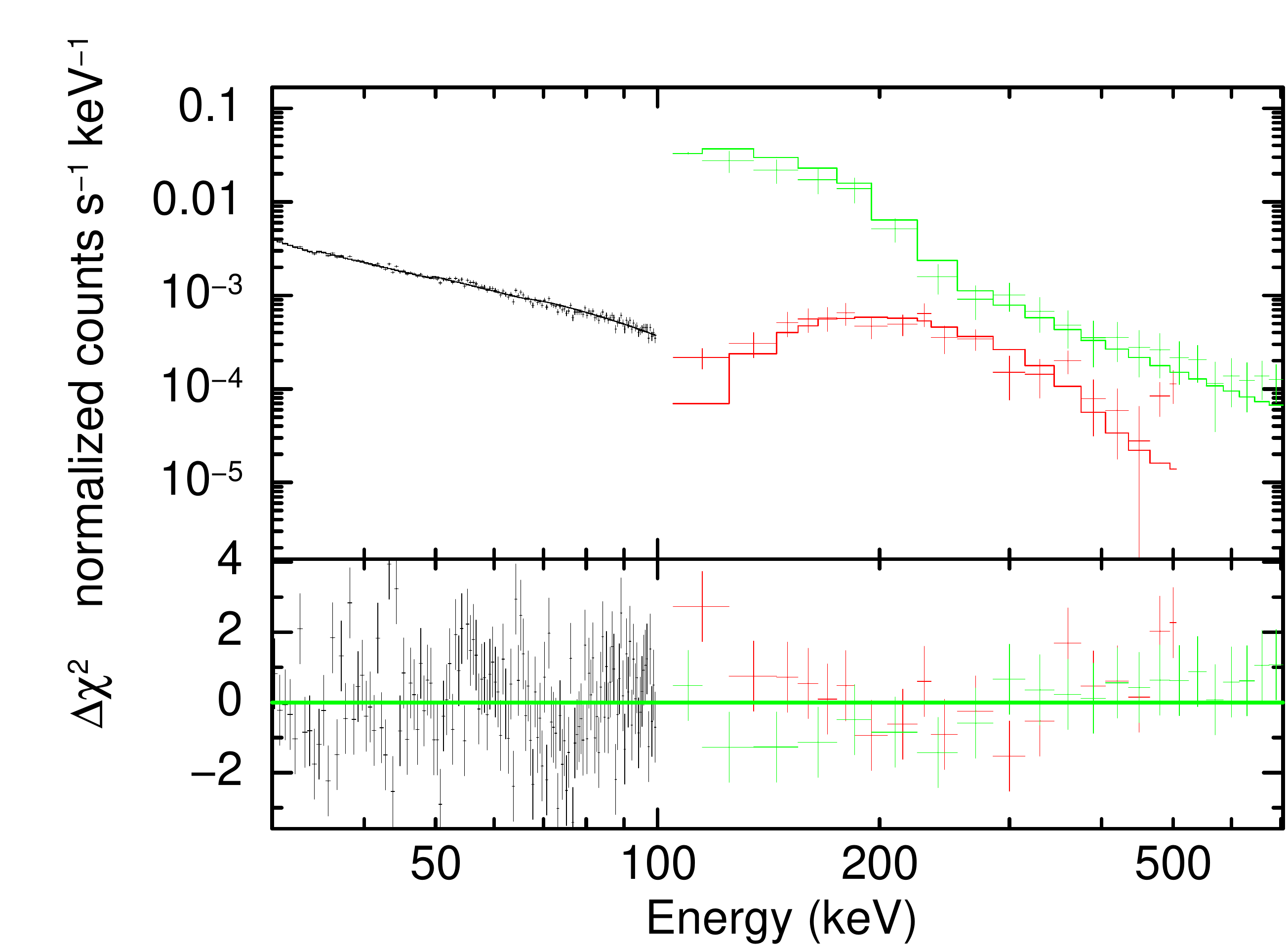}

	\caption{Broadband spectra of the Crab (ObsID 406, 114 ks) fitted with a broken power law. The black, green and red colors are used for EB1, EB2 and EB3 respectively.} 
	\label{bknspec406}
\end{figure}

\begin{figure}[!ht]
	\centering

	\includegraphics[scale=0.28]{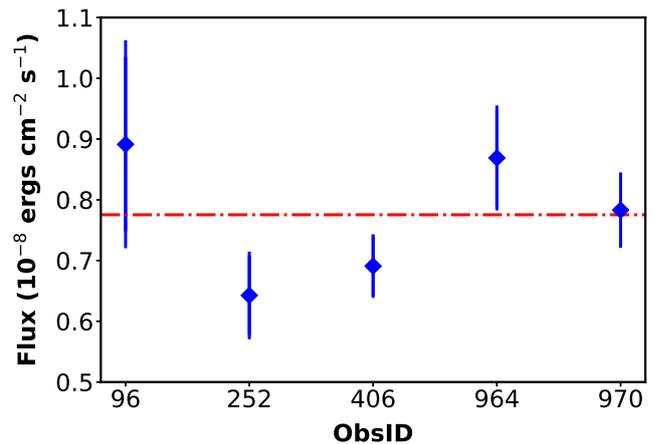}
	
	\caption{Estimated flux of the selected observations after fitting EB1, EB2, and EB3 simultaneously using broken power law. The blue diamond corresponds to Crab flux in the 100 - 700 keV energy band (EB2). The red dashed-dot horizontal lines represent the mean value ($0.78^{+0.05}_{-0.05}$) of the Crab flux of all the five selected observations added together in the corresponding energy range.} 
	\label{bknfluxEB2}
\end{figure}

 We have analysed all the selected observations and then fitted the resultant spectra (EB1, EB2, and EB3) simultaneously using the same constraints as described in section {\ref{EB1_EB3_spectra}}.  An 8$\%$ systematic error, however, have been added to EB2 to take into account of the uncertainties in the background estimation and gain calibration. The systematic error is added to the data till the residuals are uniformly distributed across zero and the spectral fit is acceptable. Spectral fitting for one of the five observations (ObsID 406, 114 ks) is shown in Figure {\ref{bknspec406}}. The fitted parameters are given in Table \ref{integral_czti_results}. The low energy slope ($PhoIndx1$) is well constrained and close to {\em INTEGRAL/SPI} (Jourdain $\&$ Roques 2008) value. The higher energy slope ($PhoIndx2$), however, is $\sim$ 38$\%$  lower than the {\em INTEGRAL/SPI} reported values, though the errors in the parameter are quite large. The influence of subtle background spectral variations and possible gain non-linearity at higher energies (the CZT detectors are not calibrated outside the energy range of 10 to 150 keV)  perhaps can lead to the somewhat flatter spectra above 100 keV. Though we are getting a flatter spectrum, we get consistent flux within 10-20$\%$ of the mean value (see Figure \ref{bknfluxEB2}). This gives enough confidence that there is a possibility of extending CZTI bandwidth up to 700 keV with better energy calibration of the low gain pixels.

\section{Discussion and conclusions}{\label{discussion}}

In this article, we have attempted to explore the sensitivity of CZTI in the sub-MeV region and have outlined a methodology of sub-MeV spectroscopy using Crab observations. For this purpose, we have used the single pixel mask-weighted spectral data in the 30 - 100 keV energy range (EB1), 2-pixel Compton events including low gain pixels in the 100 - 500 keV energy range (EB3), and the single pixel events including low gain pixels in the 100 - 700 keV energy range (EB2).   

For this work, we used the calibration parameters obtained by Chattopadhyay \emph{et al}. 2020 (in this issue) for similar work to explore the sub-MeV spectroscopic sensitivity for Gamma ray Bursts (GRBs). The advantage in the case of GRBs is the higher signal strength and, in particular, the availability of simultaneous background events before and after the burst. In the case of persistent sources, the unavailability of simultaneous background spectra makes the selection of proper blank sky flux and its subtraction extremely important (section \ref{back_sub}), particularly when the signal to noise ratio is relatively low. For this work, the spectral response was generated using a simple Gaussian energy distribution for simplicity (section \ref{response}), which we plan to improve later with the use of a more physical line profile model based on charge trapping and diffusion (Chattopadhyay \emph{et al}. 2016).

We applied these techniques for spectral analysis of the Crab, where we used a broken power law (bknpower in {\tt XSPEC}) for the spectral fitting. The spectral fits show sufficient flux sensitivity of CZTI to carry out spectroscopy for ON-axis bright sources (see Table \ref{integral_czti_results}) up to 500 keV, and it can be extended up to 700 keV with better gain calibration. We find that the overall flux measured by CZTI in 100 - 700 keV band (EB2) is consistent throughout the observations within 20$\%$ from the mean value. For the single event and Compton event spectra in 30 - 500 keV (EB1 and EB3), the low energy slope ($PhoIndx1$) and higher energy slope ($PhoIndx2$) agree reasonably with the INTEGRAL/SPI result. However, for the combined spectral fit over all three bands (EB1, EB2, EB3), the spectral parameters are not fully consistent with the results reported by INTEGRAL/SPI with
the high energy slope (PhoIndx2) showing a possible flattening. This could be either due to spectral calibration or incorrect background estimation at higher energies above 400 keV.

We note that the measured flux by CZTI is lower by $\sim$ 31$\%$ than that estimated by {\em INTEGRAL/SPI}. However, it should also be noted that in general, it is quite difficult to make a comparison of flux measurements of the two instruments, particularly in hard X-rays, due to the difficulty in measurement of the absolute effective areas of various instruments.  Considering  the errors in our flux measurements ($\sim$ 15\%) and the 5\% accuracy claimed for Crab flux measurements (Jourdain $\&$ Roques 2020), there appears to be a $\sim$ 2 sigma difference in the flux measurements of Crab made by CZTI as compared to INTEGRAL/ SPI.  We plan to do a cross-calibration with simultaneous NuSTAR observations and try to understand the effective area calibration of CZTI in a  future work.

To summarize, we find that CZTI has sufficient spectral sensitivity in the sub-MeV region for the ON-axis sources. The parameters are well constrained in the 30 - 500 keV range, but the spectra become flatter above 500 keV. 
In future, we plan to develop better background subtraction methods, and at the same time, we attempt to develop a detailed background model using 5-years of CZTI data. 
Moreover, we plan to investigate the gain of the low gain pixels in more detail; in particular, we look for various high energy background lines by proton induced radio-activation as shown by Odaka \emph{et al}. 2018. With the better characterization of the low gain pixels and background subtraction methods, CZTI is expected to provide sensitive spectroscopic information for various hard X-rays sources such as Cygnus X-1 and other bright sources and add to the better understanding of the emission mechanisms in these sources.

\section*{Acknowledgements}

This research is supported by the Physical Research Laboratory, Ahmedabad, Department of Space, Government of India. We acknowledge the ISRO Science Data Archive for {\em AstroSat} Mission, Indian Space Science Data Centre (ISSDC) located at Bylalu for providing the required data for this publication and Payload Operation Center (POC) for CZTI located at Inter University Centre for Astronomy $\&$ Astrophysics (IUCAA) at Pune for providing the data reduction software.

\end{document}